\documentstyle[preprint,aps,prl]{revtex}
\tightenlines
\newcommand{\be}{\begin{equation}} \newcommand{\ee}{\end{equation}}
\newcommand{\sst}{\scriptscriptstyle}

\newcommand{\bea}{\begin{eqnarray}} \newcommand{\eea}{\end{eqnarray}}
\def\r0{r_{\sst 0}}  \begin{document} \vskip 3.5cm

\title{Production process dependence of neutrino flavor conversion }

\author{Subhendra Mohanty}

\address{{\it Theory Group, Physical Research Laboratory, \\
Ahmedabad - 380 009, India }}

\maketitle
                                                          
\begin{abstract}

{
We perform a covariant wave-packet analysis of neutrino oscillations 
taking
into account the  lifetime of the neutrino production process . 
We find that flavor oscillations in space are
washed out when  the neutrinos are produced from long lived resonances
 - and what may be observed in appearance/disappearance experiments is a
uniform
conversion probability
independent of distance.
 The lifetime of the resonance which produces the
neutrinos acts as the the effective baseline of the experiment.
 For this reason  the LSND experiment where
neutrinos
are produced from muon decay   has two orders
of magnitude more sensitivity to neutrino mass square difference than
other experiments where the neutrinos are
 produced from pion or kaon decays.
} 
\end{abstract}

\newpage
There are two different formulas which describe flavor oscillations in
spacetime; the  one applied to  neutrino oscillations \cite{books} 
is derived in the high
energy regime 
and the other  applied to  kaon oscillations \cite{kaons}  is valid
in the non-relativistic
limit. 
From the derivations of these formulas it is not clear what formula is
 applicable at intermediate energies (for example for kaon oscillations
from stopped protons where $(m^2/E^2)\sim 25\%$) . 
A covariant
derivation of a flavor oscillation
formula which would be valid at both high and low energies 
would be of interest from the conceptual as well as the experimental point of
view.  

  Kayser and Stodolsky \cite{KS} have  advocated a                
covariant
generalization of the non-relativistic phase factor $exp (-imt)$ by
replacing the absolute time $t$ by the Lorenz invariant proper time
$s=(t^2 -x^2)^{1/2}$.  In the lab frame the covariant expression for the 
phase 
factor  is $exp -im_i s_i= exp ~(-i m_i^2 t / E_i)$. The phase difference
between
two mass eigenstates is $\Delta (m^2 /E) \simeq (\Delta m^2 /E)$ which is
twice the the phase difference of the
standard formula \cite{books}. 
This result has prompted the claim  \cite{SWS}  that
 kaon oscillations in $\Phi$ factory will have oscillation length which
is half of what is given by the standard formula. This claim has been
refuted by   
\cite{gold} who take the view that the interference
phase difference should not be evaluated at different space-time points
but should be evaluated at the average spacetime interval and taking
the
phase difference to be $(m_1-m_2) (s_1 +s_2 )/2$ instead of $(m_1 s_1 -m_2
s_2)$ they recover the
standard
formula.
Other methods of
showing that an extra factor of two does not appear in the kaon
oscillations formula have been discussed in \cite{lip2}.
A covariant derivation of the neutrino oscillation formula has been given
by Grimus and Stockinger \cite{grimus} who treat the entire process of
neutrino production ,propagation and detection as a single Feynman
diagram. They show that on taking the large distance limit of the neutrino
propagator the scattering cross section shows a space-time
oscillatory behaviour and the oscillation length is identical to that 
given by the standard formula \cite{books}. G-S assume
the initial states to be plane waves therefore the concept of a coherence
length \cite{coh} does not emerge in their formulation.

In this paper we derive the oscillation amplitude by evaluating the
Feynman propagator of the neutrinos in the large time-like asymptotic
limit. The asymptotic propagator of a position eigenstate has the form
$K (x,t; m_i) \simeq ( m_i / 2\pi i  s_i )^{3/2} ~exp{-i m_i s_i}$ where
$s_i$ is the spacetime interval propagated by the $m_i$ mass eigenstate.
This is a
field theoretic derivation of the Kayser-Stodolsky \cite{KS}
 phase factor.
 If the initial wave-functions  of the propagating particles were 
 strictly delta functions in spacetime then no interference between
different mass eigenstates can take place. We therefore
generalize the delta function propagators to propagators of Gaussian
wave-packets. The interference term as a function of distance
 is obtained by taking the time-overlap
of different mass eigenstate propagators . The
expression for flavor conversion probability (for say two flavors with
mixing angle $\theta$) as a function of distance $X$ turns out to be,
\be 
P(\alpha  \rightarrow \beta; X)= {1\over 2} sin^2 2\theta 
(1- cos({\Delta m^2 \over 2 P} X) ~e^{-A})
\label{osc1}
\ee
Therefore extra factor of two which came from naively subtracting
the
phases of 
plane wave propagators \cite{SWS} goes away in the wave-packet averaging
and the standard expression for the oscillations length is recovered. This
is in agreement with \cite{gold}-\cite{lip2}. The covariant wave-packet
treatment
introduces a new contribution to the exponential factor $A$ when particles
are produced from long lived resonances ( for example for neutrinos from
muon decay as opposed to  neutrinos from Z decay). At distances smaller
than the
coherence length , the exponential suppression factor is,
\be
A~~=~~ ( {\Delta m^2 \over 2 \sqrt{2} E}~ \tau)^2
\label{A1}
\ee 
where $\tau$ is the lifetime of the resonance which produces the
particle which undergo flavor oscillations. If 
the uncertainty in position of the
initial
particle $v \tau$  is larger than
the detector distance $X$ , the exponential
term washes out the oscillations in (\ref{osc1}). 
In  neutrino experiments
where the source is pions, kaons, muons or nuclear fission,  
the spatial oscillations of the conversion probability cannot be observed.
What can be observed in these experiments  is a constant (distance
independent) conversion
probability. 
  The conversion probability  is sensitive to 
values of $\Delta m^2 \simeq (2 \sqrt 2
 E/ \tau)$ , which means that the experimental bound on $\Delta m^2$ is
lower with longer
-lifetime sources. For this reason the
LSND
experiment \cite{LSND} which uses neutrinos from muon decay ($\tau_{\mu} =
2.19 \times 10^{-6}s$) is sensitive to two orders of magnitude lower
neutrino mass
square difference compared
to other accelerator experiments like BNL-E776 \cite{E776}, Karmen
\cite{karm} and CCFR \cite{CCFR} which use neutrinos from pion and kaon
decays ($\tau \sim 10^{-8} sec$). We fit the experimental data from LSND
along with BNL-E776, Karmen, CCFR and Bugey \cite{BUGEY} experiments with the
conversion probability formula (\ref{osc1}) and plot the allowed range
for the mass difference and mixing angle. The
lower bound on allowed $\Delta m^2$ is two orders lower than what is
obtained by fitting the LSND muon decay  data with the standard
oscillation formula
(equation (\ref{osc1}) with $A$ set to zero).

{\it  Field theoretic derivation of phase factor:}
The  phenomenon of flavor oscillations takes place because particles are
produced and
detected as weak interaction eigenstates (the neutrino states $\nu_e,
\nu_\mu$,  $\nu_\tau$ or the Kaon states $K^0 , \bar K^0$ etc)
but the propagators are diagonal in the mass eigenstates
(the neutrino mass eigenstates $\nu_i, i=1-3$ or the Kaon mass
eigenstates
$K_L , K_S$). The probability
amplitude for oscillations 
of a gauge eigenstate ($|\alpha>$) to another ($|\beta>$) is a linear
superposition
of the
propagation amplitude of the mass eigenstates ($|i>$).
\be
{\cal A}(\alpha\rightarrow \beta;t)= \sum_i~ <\beta|i>~
<i|e^{-i Ht}|i><i|\alpha>  
\label{aa}
\ee  
In relativistic field theory the propagation amplitude of the mass
eigenstates $<i|e^{-i Ht}|i>$
 can be identified with the Feynman propagator
 $S_F(x_f-x_i,m_i)= <T~\nu_i(x_i)
\bar \nu_i(x_f)>$. The Feynman propagator in position space is , 
\be
K(x,m_i) = -i \int {d^4 p \over (2\pi)^4} {(i \not \partial +
m_i)\over 
p^2 - m_i^2 -i \epsilon} e^{-i p\cdot x} .
\label{two}
\ee
This integration  can be done by  expressing the
denominator  as an exponential,
\be
{-i \over p^2 - m_i^2 -i \epsilon} = \int_0^{\infty} 
d\alpha~~
exp~\{ i \alpha ( p^2 - m_i^2 -i\epsilon)\}
\label{twoa} 
\ee
 and integrating
over the resulting Gaussian in $p$.
The remaining integral over $\alpha$,
\be
\int_0^{\infty} d \alpha ~\alpha^{-2} ~exp~ \{i \alpha m_i^2 + i{(t^2-
|{\bf x}|^2)\over 4 \alpha}\}
\label{twod}
\ee
can be performed by substituting $s=(t^2- |{\bf x}|^2)^{1/2}$, $z=i m s$
and
$\eta =2(\alpha m /s)$ and making use of the integral formula
\cite{grad} for the 
Bessel function    
\be
{1\over 2} \int_0^{\infty} d \eta ~ \eta^{-(\nu +1)}  ~exp ~
-{z\over2}(\eta+{1\over \eta}) = K_\nu (z).
\label{twoe}
\ee
The resulting expression for the Feynman propagator (\ref{two}) is
\be
K(x,m_i) = ({i\over 4 \pi^2}) (i \not \partial + m_i )~({m_i \over
s_i})~ K_1(i m_i s_i) 
\label{three}
\ee
where  $s_i= (t^2 - {\bf x}_i^2 )^{1/2}$ is the
invariant spacetime interval propagated by the $\nu_i$ mass eigenstate. If
this interval is large ($s >> m_i ^{-1}$) and time-like ( $t \ge |{\bf
x}|$ ) then we can use the asymptotic expansion of the Bessel
function 
\be
K_1(i m s) \simeq \left({2 \over \pi i m s} \right)^{1/2} ~e^{-i m s}
\label{bessel}
\ee
to obtain from (\ref{three}) the expression for the propagation amplitude
at large time-like separation
\be
K (x,t; m_i) =\left({m\over 2\pi i
\sqrt{(t_f-t)^2 -({\bf x_f} - {\bf x_i})^2)}}\right)^{3/2} ~ exp \{-im
\sqrt{(t_f-t)^2
 -({\bf x_f} - {\bf x_i}^2)} \}
\label{four} 
\ee
The phase factor of the amplitude (\ref{four}) is Lorentz-invariant and
can be
written  in terms of the neutrino energy $E$  and the time of
flight
$t$ as measured from the lab frame as
$-im_i s_i =  -i m_i~ (1-v_i ^2)^{1/2} ~t =  -i (m_i^2~ / E_i)~t$.

An
extra factor of two appears on 
subtracting the phases at different
spacetime points. A flavor eigenstates neutrino or kaon is
observed at a single spacetime point and one should therefore compute the 
phase difference at the overlap of the two mass eigenstate wave-packets.
This averaging over spacetime is done formally  by considering the
propagators of Gaussian wave-functions
as opposed to plane waves , and the standard expression for the
oscillation length is recovered.

{\it  Conservation laws for long distance propagators:}
A flavor
eigenstate propagates as a linear combination of different mass
eigenstates which are on-shell. 
The phase difference is obtained by some authors by assuming that
 the different mass eigenstates have common energy, and by some by
assuming
that they have a common momentum \cite{lip}. 
 In this
section we show that the linear combination of different mass eigenstates
have neither the same
energy nor
the same momentum. We show that 
 particles propagating  over large distances $(X >> P/m^2) $ are
constrained by the 
conservation laws at the vertex to be on shell. The on-shell condition is
all that is needed to fix the phase difference and the oscillation length.
 
Consider a diagram with a  propagator between vertices at spacetime
points $(x_1, x_2)$ with a number of external
legs at the vertices. The amplitude is proportional to
\be 
\int d^4x_1~~d^4x_2 ~~exp(-i \sum_i q_i \cdot x_1 ) ~~K(x_1,x_2)~~exp(
i\sum_f q_f \cdot x_2)
\label{c1}
\ee
Where $q_i$ are the incoming four momenta from the external legs at $x_1$
and $q_f$ are the
incoming four mommenta from the external legs at $x_2$.
Substituting the  propagator
\be
K(x_1,x_2)= \int d^4 p {e^{-i p\cdot (x_1-x_2)}\over p^2 -m^2 +i \epsilon}
\label{c2}
\ee
in (\ref{c1}) and integrating over $x_1$ and $x_2$ gives the usual energy
mommentum conservation $\delta^4 (\sum_i q_i - p)$ and $\delta^4
(\sum_f q_f - p)$ at the vertices. When the proper distance between the
two vertices $s$ is larger than $m^{-1}$, then the  propagator used in
(\ref{c1}) is
the  asymptotic form (\ref{four}), which in mommentum space can
be written
as \cite{bogol}, 
\bea
K(x_1,x_2; s>> m^{-1})&=& {1\over (2 m s)^{3/2}} e^{-i m
s}\nonumber\\[8 pt]
&=&\int {d^4 p \over \sqrt{{\bf P^2} +m^2}} ~~exp~\{-i(\sqrt{{\bf P^2}
+m^2}})(t_f-t_i)
 +i {\bf P}\cdot{\bf(x_f-x_i)\}
\label{c3} 
\eea
Substituting (\ref{c3}) in (\ref{c1}) and integrating over $x_1$ we obtain
\bea
\int d^4 x_1~ exp~ \{-i (\sum_i E_i - \sqrt{{\bf P^2} +m^2}~~)t_1 +
i(\sum_i
{\bf q_i}- {\bf P})\cdot {\bf x_1} \}\nonumber\\[8pt]
=~\delta (\sum_i E_i - \sqrt{{\bf P^2}
+m^2})~~\delta^3(\sum_i
{\bf q_i}- {\bf P}) 
\label{c4}
\eea
This implies that at a vertex
energy and
mommentum are  conserved and
in addition those particles which propagate without freely over distances
larger than $(|{\bf P}|/m^2) $ obey the mass shell constraint ,
$E= \sqrt{{\bf P}^2 + m^2}$.  

{\it  Wave-packet analysis:}
The propagator $K(x_f, x_i)$ given in (\ref{four} ) is the amplitude for
the
propagation of a
particle localized at $x_i$ (a delta function initial wave function) to
be detected at $x_f$. The propagator of some general initial 
 wave-packet $~\Psi_{in} (x-x_i)$ is obtained by
using  the expressions for the delta function propagator given in
(\ref{four}) and the
superposition principle,
\be
\tilde K(x_f-x_i)= \int d^4x ~K(x_f,x) ~\Psi_{in} (x-x_i)
\label{psif}
\ee
If 
the
initial wave function is a Gaussian ,
\bea
\Psi_{in}(x-x_i)=N~ exp \{-i E_a (t-t_i) +i {\bf P_a}\cdot ({\bf
x-x_i})\}~~
exp ~~\{ -{({\bf x-x_i})^2 \over 4
\sigma_x^2} -{(t-t_i)^2 \over 4
\sigma_t^2}  \}
\label{psii}
\eea
where $\sigma_x$ and $\sigma_t$ are the uncertainties is the initial
position and time of production of the particle respectively and $N$ is
the normalization constant.
In the
earlier wave-packet analyses of oscillation problem \cite{coh}, the
uncertainty
in the time of production was neglected. In a covariant treatment both
should be included. In the next section we show that  the time uncertainty
gives rise to a novel phenomenon of conversion without oscillations in
many experimentally relevant situations.
Substituting the expression for $K(x_f,x)$ given in
(\ref{four}) and the Gaussian initial wave-packet (\ref{psii}) in
(\ref{psif}) and evaluating the 
 integral by the stationary phase method
\cite{wolf} we obtain the expression for the  propagation amplitude
of a Gaussian wave-packet ,
\bea
\tilde K({\bf X},T;m_a)=({1\over 4 \pi |{\bf X}|^2})^{1/2}(
{\pi\over
\sigma_x^2 + v_a ^2
\sigma_t^2})^{-1/4}~~  exp\{-i{m_a^2 \over E_a} T~~
+i {\bf P_a}\cdot ({\bf X} -{\bf v_a}~T ) \} \nonumber\\[8pt]  
~\times ~exp\{ - {({\bf X} -{\bf v_a}~T  )^2\over 4( \sigma_x^2 + v_a ^2
\sigma_t^2)}\}
\label{wp}
\eea
where ${\bf X}= {\bf x_f - x_i}$ and $T=t_f-t_i$ are the space and time
intervals
propagated
by the center of the Gaussian wave-packet and ${\bf v_a}\equiv{\bf
P_a}/E_a$ . The normalization constant has been fixed such that
$\int~ 
dT~ d^3 {\bf X}~ \tilde K({\bf X},T; m_a)~~ \tilde K^{\dagger} ({\bf
X},T;m_b) = 1$.
The factor of $(1/4 \pi |{\bf X}|^2)^{1/2}$ in  (\ref{wp}) shows that the
particle flux
decreases inversely with the sqaure of distance.
Neutrino disappearance experiments  look for
evidence of depletion of a certain neutrino species over and above the
expected inverse square decrease in 
flux. In the following we will not display this factor
 $(1/ 4 \pi |{\bf X}|^2)$ in the probability expressions.
The amplitude for the oscillation of
one type of neutrino flavor to  another 
is obtained by substituting the propagation amplitudes for mass
eigenstates (\ref{wp}) in ,
\be
{\cal A}(\alpha\rightarrow \beta; X,T) = \sum_{a}^{3} U_{\alpha
a}~ \tilde K(m_a;X,T)~~~U^*_{a
\beta} 
\label{Amp}
\ee
where $\alpha, \beta $ denote the flavor eigenstates ($\nu_e, \nu_\mu,
\nu_\tau$
or $K^0 ,\bar K^0$) and the summation index $a$ denotes a mass eigenstates
($\nu_1,
\nu_2, \nu_3$ or $K_L,K_S$) . $U_{\alpha a}=<\alpha|a> $ and
$U^*_{\beta a}=
<a|\beta> $ are the
elements of
the mixing matrix which relate the flavor eigenstates with the mass
eigenstates.
The probabilty of flavor oscilation as a function of space-time is
the modulus squared of the amplitude (\ref{A}) 
\bea
P(\alpha \rightarrow \beta;X,T)&=&|\sum_{a}^{3} U_{\alpha
a}~  \tilde K(m_a;X,T)~~~U^*_{a
\beta} |^2 \nonumber\\[8pt]
&=& \sum_a |U_{\beta a}|^2~|U_{\alpha a}|^2~~+~~\sum_{a\not=b}~U_{\beta
a}~U^*_{\alpha
a}
U^*_{\beta b}~U_{\alpha b}~~\tilde K({\bf X},T; m_a)~~ \tilde K^{\dagger} ({\bf
X},T;m_b)
\label{PXT}
\eea

 In interference experiments the time of flight of the
particle is not measured and  only the
distance between the source
and
the detector is accurately known \cite{lip} . The probability for the flavor
conversion as a function of distance 
 is given by the time integral of (\ref{PXT}) ,
\bea
P(\alpha \rightarrow \beta;X)&=&\int~dT~|\sum_{a}^{3} U_{\alpha
a}~  \tilde K(m_a;X,T)~~~U^*_{a
\beta} |^2 \nonumber\\[8pt]
&=& \sum_a |U_{\beta a}|^2~|U_{\alpha a}|^2~~+~~\sum_{a\not=b}~U_{\beta
a}~U^*_{\alpha
a}
U^*_{\beta b}~U_{\alpha b}~~\int~ 
dT \tilde K({\bf X},T; m_a)~~ \tilde K^{\dagger} ({\bf
X},T;m_b)
\label{PX}
\eea
The interference term  given by the
time-overlap of the propagation amplitudes (wave-functions) of different
mass eigenstates can be evaluated for the Gaussian propagator (\ref{wp})
and is given by,
\bea
 Re ~\int dT~~\tilde K({\bf X},T; m_a)~~ \tilde K^{\dagger} ({\bf
X},T;m_b)
&=& ~
~{2\over v}~ cos
\{ ({\bf P_a -P_b})\cdot {\bf X} - (E_a -E_b){({\bf v_a +
v_b})\over 2 v^2}\cdot {\bf X} \}\nonumber\\[8pt]
~ \times exp\{&-&(E_a-E_b)^2 {( \sigma_x^2 +
 v^2\sigma_t^2)\over 2  {v^2}}- {({\bf v_a - v_b})^2 ~X^2
\over 8
 ( \sigma_x^2 +   {v^2} \sigma_t^2)   {v^2}} \}    
\label{intX}
\eea
where  $  {v}\equiv \sqrt{(v_a^2 + v_b^2)/2}$.
 In terms of the average momentum ${\bf P}$ and
energy $E$
and their respective differences $\Delta {\bf P}$and
$\Delta E$ the interference term (\ref{intX}) ,to the
leading order in
$(\Delta P/ P)$ and $(\Delta E/ E)$, reduces to the form,
\be
~{2\over v}~ cos
\{{X\over  P}( E \Delta E - {\bf P} \cdot \Delta {\bf
P})\}~exp\{ -A \} 
\label{I}
\ee
where the exponential damping factor,
\be
A=(\Delta E)^2~  {\bar \sigma^2~\over 2}({ E^2\over
P^2})+ ({\Delta P\over P})^2
~{X^2 \over 4  \bar \sigma^2}    
\label{AA}
\ee
and $\bar \sigma^2 \equiv ( \sigma_x^2 + (P/E)^2 \sigma_t^2)$.

In general neither $\Delta E$ nor 
$\Delta P$ is zero and they depend upon how the state is prepared.
For example if the mommentum of the initial and the associated final state
is measured then $\Delta P=0$. 
 In the last section we have shown
that for long distance propagation ($s > m^{-1}$)
 ${\bf P}$ and $E$  are not
independent and are related by the mass
shell constraint.
The particular combination that appears in the phase difference
 turns out to be independent of the
preparation and is fixed by the condition that each of the mass
eigenstates be on shell,
\bea
{E_i}^2 -{P_i}^2 &=& {m_i}^2   ~~~~~~~~i= a,b \nonumber\\[8pt]
\Rightarrow~~ 2 E \Delta E -2 {\bf P}\cdot  {\bf \Delta P} &=& \Delta m^2
\label{mshell}
\eea
Substituting (\ref{mshell} ) in  (\ref{I}) we see that the
interference 
term of  (\ref{intX}) is given by ,
\be 
{2\over v}~cos ( {\Delta m^2 \over 2 P} X) ~~e^{-A}~~~~~~~~~~,  
\label{cosA}
\ee
The probability for the flavor
conversion as a function of distance (\ref{PX}) is therefore,
\bea
P(\alpha \rightarrow \beta;X)&=&
 \sum_a~{1\over v_a}~ |U_{\beta a}|^2~|U_{\alpha
a}|^2~~\nonumber\\[8pt]
&+&~~\sum_{a\not=b}~ {1\over v}~~|U_{\beta
a}~U^*_{\alpha
a}
U^*_{\beta b}~U_{\alpha b} | cos( {(m_a^2 -m_b^2)\over 2  P}X
~~+~~\delta)~~e^{-A}
\label{covP}
\eea
where  $~\delta
=arg(U_{\beta a}~U^*_{\alpha a}
U^*_{\beta b}~U_{\alpha b} )$.

  We see that the
 standard oscillation formula which was obtained for relativistic
particles \cite{books} is actually valid at all energies.
In other words although the formula (\ref{covP}) is usually derived by
taking the leading term in a $(m^2/P^2)$ series, our covariant calculation
shows that there are actually no $O(m^2/P^2)$ corrections to (\ref{covP}). 

 In
the non-relativistic regime where $P=(m_a+m_b) v /2 $, the standard kaon
oscillation formula $ cos (m_a-m_b) T $ is recovered.

There is no extra
factor
of two in the relativistic kaon oscillation 
formula  as would have appeared
without the wavepacket averaging .  We see that although
the two mass eigenstates have different proper times the wave packet
overlap results in the average proper time appearing in the interference
term. Therefore the interfernce phase is actually $(m_a-m_b) \bar{s}
=(m_a-m_b)
(m_a+m_b) T /(E_a +E_b)= (m_a^2-m_b^2) T/2E $. The average
time prescription of \cite{gold}
 can therefore be
justified using the covariant propagator method. Other methods of
showing that an extra factor of two does not appear in the flavor
oscillations formula have been discussed in \cite{lip2}.

{\it  Conversion without oscillation:}
The suppression factor $ A$ which goes with the oscillation term has
some
interesting new implications. If the parameters of the experiment are such
that $A$ becomes large then no oscillations is spacetime can be observed.
What can be observed is a constant conversion probability.
 
It has been noted  ealier \cite{coh}
one condition $A$ must be small
which implies that $X $ must be smaller than the coherence  length
$L_{coh}=(2 \bar \sigma P/\Delta P)$ .
      We shall show below that the
constraint $L_{osc} < L_{coh}$ is not a sufficient condition to ensure the
occurrence of flavor ocillations in space.
In the 
analysis of refs \cite{coh} only the uncertainty in the
initial position $\sigma_x$ was considered,
 in our analysis we have
included the contribution of $\sigma_t$ the uncertainty in time at which
the neutrino was produced , which as it turns out, makes the larger
contribution to the
suppression term $A$. This happens when the neutrinos are produced from
long lived resonances. The neutrino wavepacket which is produced has a
spread in time with width $\sigma_t$ which cannot be smaller than the
lifetime of the resonance whose decay produces it.
The dominant contribution in that case arises from 
the first term of $A$,
\be
A\simeq (\Delta E)^2~{\bar \sigma^2 \over 2} \simeq ({\Delta m^2
~\tau \over 2 \sqrt{ 2} E})^2
\label{A}
\ee
where we have equated the initial time uncertainity $\sigma_t$ with
$\tau$ - the
lifetime of the resonance that produces the neutrino, and we have ignored
the spatial spread of the wave-packet $\sigma_x$ which in most
experiments is many  orders of magnitude smaller than $\sigma_t$. 
In terms of the oscillation length $ L_{osc}= 4 \pi E/ (\Delta m^2)$
the suppression factor $A= ({\sqrt 2} \pi \tau/ L_{osc})^2 $. 
In order to observe oscillations the  source-detector distance
$L$ must be larger than $L_{osc}$.
Which  that the
minimum source-detector distance in order that neutrino oscillations be
observed is given by condition,
 $~L_{min}~=~(\pi)\sqrt{2} \tau $.
We shall show below that most neutrino experiments do not satisfy this
criterion for observability of space oscillations. On the other hand
spatial oscillations of flavor  can be observed in the $\Phi$  and $B$
factories owing to the large width of $\phi$ and $\Upsilon$ resonances.

For relativistic particles produced from long lived resonances the 
formula for the conversion probability 
which should be fitted with the experiments is ,
\be
P(\nu_{\alpha} \rightarrow \nu_{\beta})={1\over2}~sin^2 2\theta ~
~~\left(1-~cos({2.53 \Delta m^2 L \over E})~~exp\{-({1.79 \Delta m^2 \tau
\over
E})^2\} ~\right)
\label{cor}
\ee 
where $\Delta m^2$ is the mass square difference in $eV^2$ ,$L$ is the
detector distance in $m~(km)$,  $\tau$ is the lifetime of the parent
particle
in the lab frame
in $m~(km)$ and $E$ is the  energy
in $MeV~(GeV)$. The limits on the values of $\Delta m^2$ and $sin^2
2\theta$ obtained by fitting the results of different experiments with
the oscillation formula (\ref{cor}) are listed in Table I.

\begin{table}\squeezetable
\caption{The asymptotic limits on $\Delta m^2$ and
 $sin^2 2
\theta$
from different experiments according to the
oscillation formula (\ref{cor}). $\tau$ is the lifetime of the neutrino
source in the lab frame, $<E_{\nu}>$ is the average $\nu$ energy, $L$ is
the detector
distance, $L_{min}$ is the minimum detector distance for observing
oscillations in space and $P$ is the experimental value of the conversion
probability.} 

\label{Sig} \begin{tabular}{|c|c|c|c|c|c|c|c|}
 Experiment(Source) &$\tau~(m) $  &$<E_{\nu}>~(MeV)$& $L~(m)$
&$L_{min}~(m)$&$P$ &$\Delta m^2~(eV^2)$ &$ sin^2
2 \theta$ \\
\tableline
LSND ($\mu$)  \cite{LSND}& $658.6$  &$ 30 $&$ 30 $& $ 2927$ &$(0.16 -
0.47)\times 10^{-2}$& $(2.0-3.0 )\times 10^{-3}$& $0.003 - 0.009$ \\
\tableline
LSND ($\pi$)\cite{lsnd2}& $17$&$130$&$30$& $76 $ & $(0.26 \pm
0.15)\times 10^{-2}$&
$0.8-1.6 $& $0.002-0.0082$\\  
\tableline
Karmen ($\pi$)\cite{karm}& $7.8$&$29.8$&$17.5$& $34.6$ & $< 0.3
\times
10^{-2}$&
$< 0.16$& $< 0.6 \times 10^{-2}$\\
\tableline
E776 ($\pi$) \cite{E776}& $578$&$5\times10^3$&$10^3$& $2.6\times10^3$ 
& $< 0.15\times 10^{-2}$&$< 0.2$& $< 0.3\times 10^{-2}$\\  
\tableline
CCFR( $K$)\cite{CCFR}& $5.41 \times 10^3$ & $140 \times 10^3$ & $1.4
\times
10^3$&$24\times 10^3$&
$<0.9 \times 10^{-3}$& $<1.2$ & $< 0.18 \times 10^{-3}$\\  
\tableline
Bugey(U,Pu) \cite{BUGEY}& $3 \times 10^{10}$ & $5$& $95$ &
$ 10^{11}$ & $ < 0.75 \times 10^{-1}$ & $< 10^{-9} $ & $<0.15$\\
\tableline
\end{tabular}
\end{table}

\begin{figure}
\vskip 20cm
\includegraphics{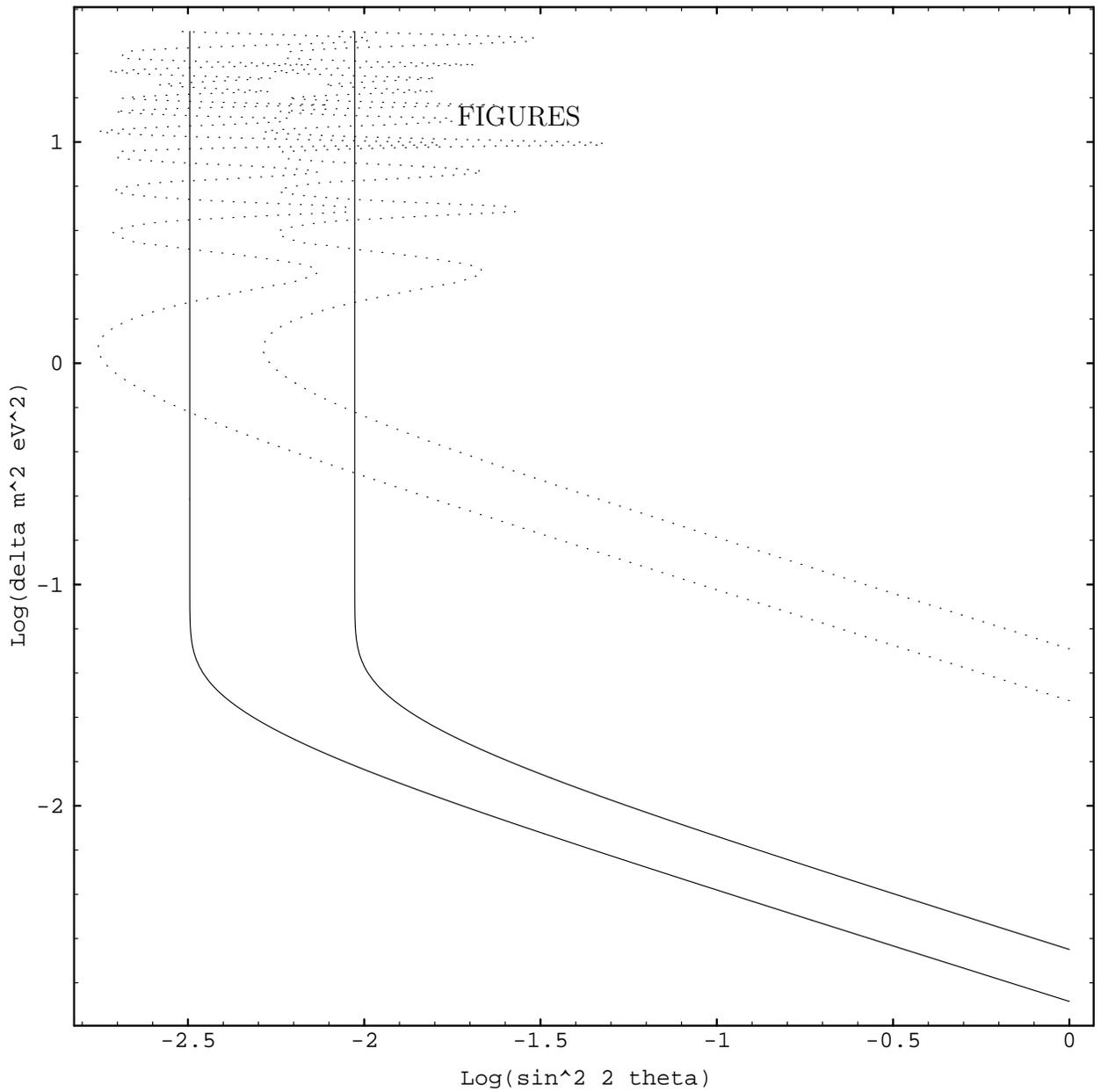}
\caption[dummy]{ Lsnd allowed regions with the standard (between the
dotted lines) and the covariant
oscillation formulas (between the continuous line). The conversion
probability formulae are averaged
over a
Gaussian distribution of neutrino energy with mean $30 MeV$ and width
$10 MeV$. }
\label{fg:lsnd}
\end{figure}
\newpage
.
\begin{figure}
\vskip 20 cm
\includegraphics{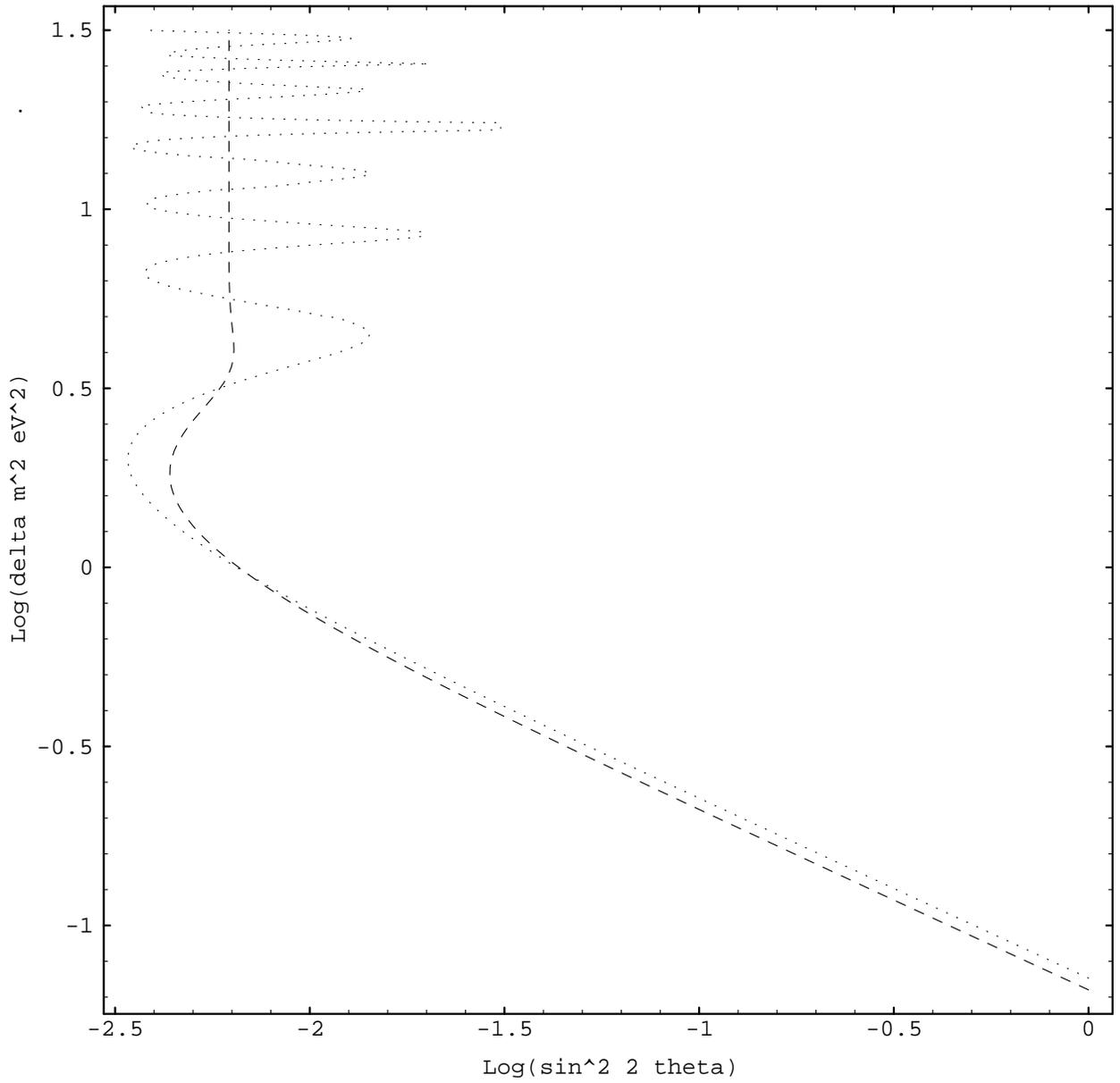}
\caption[dummy]{ Karmen allowed regions with the standard (dotted curve) 
and the
covariant oscillation formulas (dashed curve).The conversion probability
formulae are averaged
over a
Gaussian distribution of neutrino energy with mean $30 MeV$ and width
$10 MeV$. }
\label{fg:karm}
\end{figure}
\newpage
.
\begin{figure}
\vskip 20 cm
\includegraphics{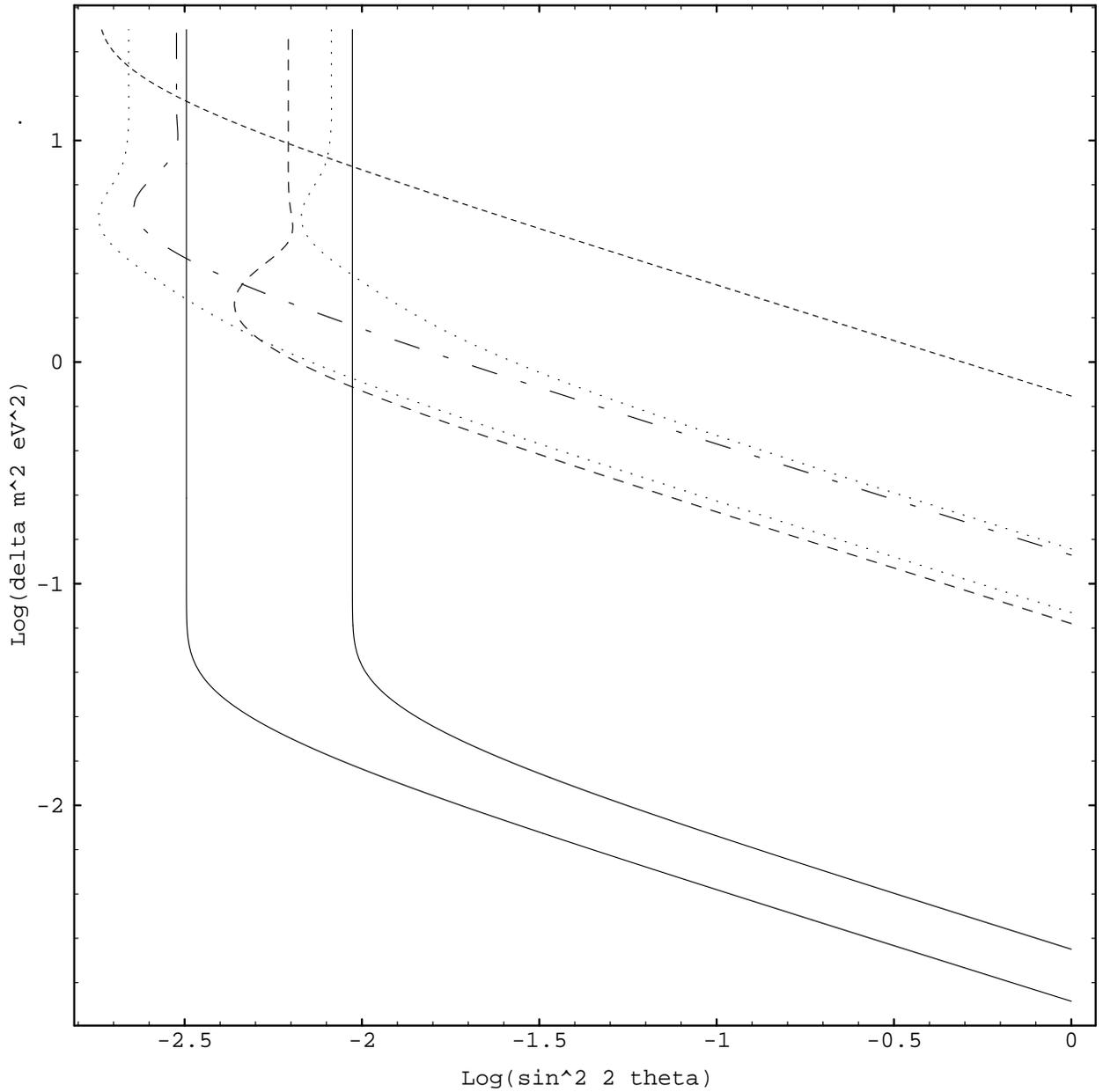}
\caption[dummy]{The combined fit of the covariant oscillation formula
with all experiments.The region between the continuous lines is allowed by
the LSND $\mu$
experiment \cite{LSND} .The region between the dotted
lines is
allowed by the
LSND $\pi$ experiment \cite{LSNDpi} . Region ruled out
by E776 \cite{E776} is above the dashed-dotted curve and by Karmen
\cite{karm} is above dashed
curve. The region above the top-most dashed curved is ruled out by
CCFR \cite{CCFR}.
}
\end{figure}

\newpage

In the standard oscillation formula the oscillatory term averages to zero 
 when $\Delta m^2 >(4\pi E/L)$ . Using the
wavepacket
formula (\ref{cor}) however we see that the oscillatory term is 
is exponentially damped at much lower values of $\Delta m^2$  (provided
$\tau >(\sqrt 2 2 L/4 \pi)$).
If neutrino sources have a large decay time  which is
the case in most experiments , the conversion probabilty is sensitive to
mass differences  $\Delta m^2 > (\sqrt 2 2 E/\tau)$. 
For this reason the LSND
experiment which uses neutrinos from muon decay has two orders of
magnitude more sensitive than  KARMEN, BNL-E776 etc  where the neutrinos
are from $\pi$ and $K$ decay. 
This is evident by comparing the fit of LSND $\mu$ DAR allowed region Fig1
 with the
Karmen allowed region Fig 2.
In experiments where the source is in flight w.r.t the detector the
gain in the source lifetime by the Lorentz factor $\gamma =
(E_{\pi}/m_{\pi})$ is somewhat compensated by the larger $E_{\nu}$ in the
denominator of $A$ (\ref{A}), so there is no gain in sensitivity with
in flight neutrino sources (as in CCFR and BNL-E776 experiments).
The reactor experiments where the neutrinos arise from the
$\beta$-decay of $^{235} U$ and $^{229}Pu$ nuclei , $\tau \sim
10^2 s$ and therefore these experiments are in fact
sensitive to values of $\Delta m^2 $ as low as $10^{-9} eV^2$. But the
probability measurement in reactor experiments is poor compared to the
accelerator experiments
which is why they rule out only a small region of parameter space.  
If the recent LSND $\pi$ decay in flight result \cite{lsnd2} is taken as
an actual 
observation of $\nu_{\mu} -> \nu_e $ conversion then it rules out most of
the low $\Delta m^2$ region allowed by the LSND $\mu$ decay experiment.

{\it Conclusions}

The covariant formulation gives a  very different result for the
oscillation probability formula compared to the standard
treatment
when the neutrino source has a lifetime larger than the baseline of the
experiment. We use the condition that the minimum spread of the neutrino
wave function along the time axis is given by the lifetime of the particle
whose decay produces the neutrino. When neutrinos are produced from long
lived particles like muons , the interference term in the conversion
probability formula vanishes for much smaller values of the
 neutrino
mass square difference compared with the standard formula. The reason
this happens can be understood with the following picture. When the spread
of the neutrino is large along the time axis, , its spread along the energy
axis is small. The neutrino wave-functions for different masses are then
energy eigenstates which are mutually orthogonal. The interference term 
which is the overlap of the wave-functions of different  mass states
therefore vanishes when these states become orthogonal.

{\it Acknowledgments} I thank Terry Goldman, Walter Grimus and Marek
Nowakowski for useful correspondence.


\begin{references}                                

\bibitem{books} B. Pontecorvo, Sov. Phys. JETP {\bf 26}, 984 (1968);\\
              S. M. Bilenky and S. T. Petcov, Rev. Mod. Phys. {\bf 59},
              671 (1987);\\
              R. N. Mohapatra and P. B. Pal, {\it Massive Neutrinos in Physics
              and Astrophysics} (World Scientific, Singapore, 1991);\\
              C. W. Kim and A. Pevsner, {\it Neutrinos in
              Physics and Astrophysics} ,harwood academic publishers, 
              Chur,(1993).                              
\bibitem{kaons} E.D.Cummins and P.H.Bucksbaum, {\it Weak interactions of
leptons and quarks}, Cambridge University Press,(1983);\\
R.P.Feynman , R.B.Leighton and M.Sands 
{\it Feynman Lectures in Physics Vol III}, 
Addison Wesley, Reading  MA (1965).                            
\bibitem{KS}  Kayser and Stodolsky, Phys. Lett.{\bf  B359}, 333(1995). 
\bibitem{SWS} Y.N.Srivastava, A.Widom and E.Sassaroli, Z. fur Physik,
{\bf C66},601 (1995).                 
\bibitem{gold} J.Lowe, B. Bassallek, H. Burkhardt, A.Rusek, G.J.Stephenson Jr
and T.Goldman , Phys. Lett {\bf B384}, 228(1996) ;\\  
B.Kaysor, hep-ph 9702327.   
\bibitem{lip2} Yu. Grossman and H.J.Lipkin, Phys Rev {\bf D55}, 2760 (1996);\\
B.Anchoea, A.Bramon, R.Muniz-Tapia and M. Nowakowski ,
Phys Lett {\bf B389},\\149(1996) ;\\
A.Dolgov, A.Yu.Morozov, L.B.Okun and M.G.Schepkin hep-ph 9703241;\\
M.M. Nieto, hep-ph 9509370 ;\\
T. Goldman , hep-ph 9604357.
\bibitem{grimus} W.Grimus and P. Stockinger, Phys Rev {\bf D54}, 3414 ,(1996).
\bibitem{coh} C. Giunti, C. W. Kim, J. A. Lee and U. W. Lee,
              Phys. Rev. D {\bf 48}, 4310 (1993);\\
              B. Kayser, Phys. Rev. D {\bf 24}, 110 (1981);\\
              S.Nussinov, Phys Lett{\bf B63} , 201 (1976).
\bibitem{LSND}  C.Athanassopoulos et al (LSND ) , Phys Rev Lett. {\bf 75} 
, 2650 (1995).
\bibitem{E776} L.Brodovsky et al (BNL-E776), Phys. Rev Lett {\bf 68}, 274 (1992 ).
\bibitem{karm} B.Armbrusuter et al (Karmen), Nucl.Phys. {\bf B38 },235 (1995).
\bibitem{CCFR} A. Romosan et al (CCFR), Phys. Rev. Letters  ,{\bf 78}, 2912(1997).
\bibitem{BUGEY} B.Achkar et al (Bugey), Nucl. Phys. {\bf B434} , 503 (1995).
 \bibitem{grad} I.S.Gradshteyn and I.M.Ryzhik, Academic Press, New York,
              p 956 , 8.427, (1981).
\bibitem{lip} H.J. Lipkin , Phys. Lett.{\bf B348}, 604 (1995).     
\bibitem{bogol} N.N. Bogoliubov and D.V. Shirkov, {\it An introduction to
the theory of quantised fields }, Wiley Interscience, New
              York, (1959). 
\bibitem{wolf} L. Mandel and E.Wolf, {\it Optical coherence and quantum
optics},
p 128, Cambridge University Press, (1995).
\bibitem{lsnd2} C.Athanassopoulos et al (LSND ) , nucl-ex/9706006. 
\end{references}
\end{document}